\documentclass[prd,preprint,superscriptaddress,amsmath,amssymb,nofootinbib]{revtex4}
\usepackage{graphicx}
\usepackage{dcolumn}
\usepackage{bm}
\usepackage{amssymb}
\usepackage{amsmath}
\usepackage{epsfig}    
\usepackage{color}
\usepackage{slashed}
\usepackage{hhline}

\def\be{\begin{equation}}
\def\ee{\end{equation}}
\newcommand{\bea}{\begin{eqnarray}}
\newcommand{\eea}{\end{eqnarray}}
\newcommand{\nn}{\nonumber}

\begin{document}

{\begin{flushright}{APCTP Pre2020-020}
\end{flushright}}

\title{Neutrino and dark matter in a gauged $U(1)_R$ symmetry} 
\author{Keiko I. Nagao}
\email{nagao@dap.ous.ac.jp}
\affiliation{Okayama University of Science, Ridaicho 1-1, Kita-ku, Okayama, 700-0005, Japan}
\author{Hiroshi Okada}
\email{hiroshi.okada@apctp.org}
\affiliation{Asia Pacific Center for Theoretical Physics (APCTP) - Headquarters San 31, Hyoja-dong,
Nam-gu, Pohang 790-784, Korea}
\affiliation{Department of Physics, Pohang University of Science and Technology, Pohang 37673, Republic of Korea}

\date{\today}

\begin{abstract}
We study neutrinos and dark matter based on a gauged $U(1)_R$ symmetry in the framework of a radiative seesaw scenario.
We identify the dark matter as a bosonic particle that interacts with the quark and the lepton sectors through vector-like heavier quarks and leptons.
The dark matter also plays a role in generating the neutrino mass matrix with the neutral heavier fermions.
We explore several constraints for the masses and the couplings related to the dark matter by computing the relic density and the scattering cross sections for direct detection methods, taking into consideration neutrino oscillations, lepton flavor violations, and the muon anomalous magnetic moment.
Finally, we mention the semileptonic decays and the neutral meson mixings that occur through the dark matter by one-loop box diagrams.
\end{abstract}
\maketitle

\section{Introduction}
It is important to understand the neutrino sector, which is typically discussed in beyond the standard model (SM) scenarios.
The simplest way to construct the neutrino mass matrix is to introduce heavier right-handed neutrinos in a renormalizable theory.
If one needs a principle to introduce them, a gauged Baryon minus Lepton number symmetry $U(1)_{B-L}$~\cite{Mohapatra:1980qe, Davidson:1978pm} or a right-handed symmetry $U(1)_{R}$~\cite{Jung:2009jz, Ko:2013zsa, Nomura:2016emz,Nomura:2017tih,Chao:2017rwv, Nomura:2018mwr, Nomura:2017ezy, Nomura:2016pgg, Nomura:2017lsn,Jana:2019mez,Seto:2020jal,Seto:2020udg} are promising candidates. 
They are promising because both these symmetries demand three families of right-handed neutrinos, so as to cancel chiral anomalies.
In addition, the $U(1)_R$ symmetry requires zero new charges for the left-handed fermions, and leads to richer phenomenologies than the $U(1)_{B-L}$ symmetry.
The current Large Hadron Collider(LHC) constraint on the $Z'_R$ mass is stronger(weaker) than that from $v'$ for $m_{Z'_R} \lesssim(\gtrsim) \ 3.9$ TeV, and a wider parameter region of $\{g', m_{Z'_R} \}$ can be tested by the future LHC experiments~\cite{Nomura:2017tih}, where $\{v', g', m_{Z'_R} \}$ are respectively the gauged $U(1)_R$ vacuum expectation value(VEV), coupling, and a neutral vector boson.
The chirality structure of $U(1)_R$ could be tested at the LHC, by measuring forward-backward and top polarization asymmetries in the $Z'_R \to t \bar t$ mode, which can distinguish the $Z'_R$ interaction from the other $Z'$ interactions like $U(1)_{B-L}$.
The effective interaction $(\bar e_R\gamma^\mu P_R e)(\bar f_{\rm SM} \gamma_\mu P_R f_{\rm SM})$ can also be tested by measuring the process $e^+ e^- \to f_{\rm SM} \bar f_{\rm SM}$ at the International Linear Collider (ILC)~\cite{Baer:2013cma}, even if the $Z'_R$ mass is too heavy to be directly produced at the LHC. 
In particular, analysis of the polarized initial state is useful to distinguish our model from other $Z'$ models, such as the $U(1)_{B-L}$ models, since $Z'_R$ only couples to the right-handed SM fermions.


It is also essential to involve a non-baryonic dark matter (DM) candidate 
into the SM model, which occupies 26.5\% of the Universe~\cite{PDG2019}. Weakly Interacting Massive Particles (WIMPs) are one of the promising candidates for DM to explain the observed relic density in the Universe.
In direct detection searches, the {\it smoking gun} for DM has not been obtained yet, but we have a common understanding that stringent constraints are imposed on spin independent scattering cross sections of DM~\cite{Schumann:2019eaa,Aghanim:2018eyx}.
It is worthwhile to mention that the direct detection search has shifted to the lighter DM mass regions~\cite{Ren:2018gyx,Agnes:2018ves,Abdelhameed:2019hmk,Aprile:2018dbl, Aalbers:2016jon,Agnese:2016cpb,Arnaud:2017bjh,Akerib:2018lyp}, and directional sensitivity is also being taken into consideration~\cite{Ahlenet:2010,Mayet:2016zxu,Aleksandrov:2016fyr,Yakabe:2020rua,OHare,Kavanagh,Nagao:2017yil}.
%
%
 
In this paper, we consider neutrino and WIMP-DM particles based on the gauged $U(1)_R$ symmetry in a framework of the radiative seesaw scenario~\cite{Ma:2006km}.
This radiative seesaw scenario is renowned as an elegant method to connect the neutrino sector and DM sector at loop levels.
In other words, the neutrinos do not couple to the SM Higgs in a direct manner.
To realize our radiative seesaw model, we introduce an additional $Z_2$ symmetry
to forbid the Dirac neutrino mass term at the tree level.
We also introduce vector-like fermions, which are automatically anomaly free, additional neutral fermions with zero-charges under the $U(1)_R$ symmetry, and several bosons.
These vector-like fermions play a role in connecting the DM and the SM fermions; in addition, they provide fruitful phenomenologies for the DM physics and the neutrino sector.

This letter is organized as follows.
In Sec. II, we introduce our model by writing down all the allowed terms for our Lagrangian.
Then, we construct the neutral fermion mass matrix including the active neutrino mass matrix, and calculate the lepton flavor violations (LFVs) and the muon anomalous magnetic moment (muon g-2). We estimate the muon g-2 in terms of the relevant Yukawa couplings.
In Sec.III, we discuss the DM content and estimate its mass and the valid Yukawa couplings by considering constraints from the relic density and the direct detection searches.
In Sect.IV, we discuss the other phenomenological possibilities in our model such as the semi-leptonic decays and the neutral meson mixings.
Finally, we devote the last section to a summary of our results and the conclusion.

\section{Model setup and Constraints}
\begin{table}[t!]
\begin{tabular}{|c||c|c|c|c|c|c|c|c|c||c|c|c|}\hline\hline  
& ~$Q_L^a$~& ~$u_R^a$~  & ~$d_R^a$~ & ~$Q'^a$~& ~$L_L^a$~& ~$e_R^a$~& ~$N_R^a$~& ~$S_L^a$~& ~$L'^a$~& ~$H(H')$~ & ~$\varphi$~& ~$\chi$~\\\hline\hline 
$SU(3)_C$ & $\bm{3}$  & $\bm{3}$ & $\bm{3}$& $\bm{3}$ & $\bm{1}$ & $\bm{1}$ & $\bm{1}$ & $\bm{1}$ & $\bm{1}$ & $\bm{1}$ & $\bm{1}$ & $\bm{1}$   \\\hline 
$SU(2)_L$ & $\bm{2}$  & $\bm{1}$  & $\bm{1}$  & $\bm{2}$& $\bm{2}$  & $\bm{1}$  & $\bm{1}$  & $\bm{1}$ & $\bm{2}$   & $\bm{2}$   & $\bm{1}$ & $\bm{1}$   \\\hline 
$U(1)_Y$   & $\frac16$ & $\frac23$ & $-\frac13$ & $\frac16$ &  $-\frac12$  & $-1$ & $0$  & $0$  & -$\frac12$ & $\frac12$ & $0$  & $0$\\\hline
$U(1)_{R}$   & $0$ & $x$ & $-x$ & $x$  & $0$  & $-x$  & $x$  & $0$  & $x$ & $x(-x)$   & $2x$  & $x$\\\hline
$Z_2$   & $+$ & $+$ & $+$ & $-$  & $+$  & $+$  & $-$  & $-$  & $-$ & $+$  & $+$  & $-$ \\\hline
\end{tabular}
\caption{ 
Charge assignments of the our fields
under $SU(3)_C\times SU(2)_L\times U(1)_Y\times U(1)_{R}\times Z_2$, where the upper index $a$ is the family number that runs over 1-3, and $H'$ has the same charges as $H$ under these symmetries except $U(1)_R$ symmetry.}
\label{tab:1}
\end{table}

In this section, we formulate our model.
To begin with, we introduce three families of right(left)-handed isospin singlet fermions $N_R(S_L)$ with $x(0)$ charge under the $U(1)_R$ gauge symmetry,
and the isospin doublet vector fermions $Q'\equiv[u',d']^T$ and $L'\equiv[N',E']^T$ with a $x$ $U(1)_R$ charge. 
\footnote{Although we introduce three families of $L',Q'$ for simplicity, three families are not needed.
In the case of $L'$, two families are enough in order to reproduce neutrino oscillation data. In the case of $Q'$, one family may be fine because it does not affect any masses of SM. }
Here, $x$ is an arbitrary nonzero number.
Next, we include an isospin singlet boson $\varphi$ with a $2x$ $U(1)_R$ charge, an inert boson $\chi\equiv (\chi_R+i\chi_I)/\sqrt2$
with a $x$ $U(1)_R$ charge, and an isospin doublet $H_{1}$ with a $x$ $U(1)_R$ charge.
We denote each of the vacuum expectation values to be $\langle H\rangle\equiv [0,v_{H}/\sqrt2]^T$,  $\langle H'\rangle\equiv [0,v_{H'}/\sqrt2]^T$, and $\langle \varphi\rangle\equiv v_{\varphi}/\sqrt2$.
Furthermore, the SM Higgs boson $H$ also has a $x$ $U(1)_R$ charge to induce all the masses of the SM fermions 
after spontaneous symmetry breaking. While $H'$ has the same charges as $H$, except the $U(1)_R$ symmetry,
and $H'$ plays a role in connecting $L'$ and $S$ that leads to the neutrino mass matrix at the one-loop level as shown in Fig.~\ref{fig:neutrinomass}.
All the field contents and their assignments are summarized in Table~\ref{tab:1}.
The relevant Yukawa Lagrangian and Higgs potential under these symmetries are given by:
\begin{align}
-{\cal L_\ell}
&= y_u^{ii}\bar Q^i\tilde H u_R^i + y_d^{ij}\bar Q^i  H d_R^j
+  y_{\ell_{ii}} \bar L^i_L H e^i_R  +  f_{\alpha i} \bar L'^\alpha_R L^i_{L}\chi  +  g_{ab} \bar L'^{a}_R  S^{Cb}_{L} \tilde H'
  + h_{\alpha i} \bar Q'^\alpha_R Q^i_L\chi \nn\\
&+y_{N_{aa}} \varphi^*\bar N^a_R N^{c a}_R + M_{S_{aa}} \bar S^a_L S^{c a}_L + M_{L'_{\alpha\beta}} \bar L'^{\alpha}_L L'^\beta_R+ M_{Q'_{\alpha}} \bar Q'^{\alpha}_L Q'^\alpha_R
+ {\rm h.c.} \label{Eq:yuk}, \\
 {\cal V} &= {\cal V}_{2}^{tri} + {\cal V}_4^{tri} + \mu_1\varphi^* \chi^2+\mu_2\varphi H^\dag H'
 + {\rm h.c.}
,\label{Eq:pot}
\end{align}
where $\tilde H\equiv i\sigma_2H^*$, upper indices $i,j,\alpha,a,b=1-3$ are the flavors number, and
$y_u, y_\ell,y_N$, $M_S, M_{Q'}$ are assumed to be diagonal.
The $\mu_1$ term plays a role in generating the mass difference between $\chi_R$ and $\chi_I$ given as,
$m_R^2-m_I^2=\sqrt2 \mu_1 v_\varphi$; where $m_{R,I}$ are the mass eigenstates of $\chi_{R,I}$.
While the $\mu_2$ term forbids a dangerous massless boson originating from $H'$.
Lastly, the ${\cal V}_{2}^{tri}$ and ${\cal V}_4^{tri}$ are trivial quadratic and quartic terms of the Higgs potential, respectively; therefore, 
$
{\cal V}_{2}^{tri}= \sum_{\phi=H,H', \varphi,\chi} \mu_{\phi}^2|\phi|^2,\text{ and }
{\cal V}_{4}^{tri}= \sum_{\phi'\le\phi}^{H,H',\varphi,\chi} \lambda_{\phi\phi'} |\phi|^2|\phi'|^2 +\lambda'_{HH'} |H^\dag H'|^2.\label{Eq:pot-tri}
$

\subsection{Neutral fermions}
After the spontaneous symmetry breaking, the 9$\times$9 neutral fermion mass matrix that is based on $[N'_R,N'^c_L,S^c_L]^T$ is given by
\begin{align}
M_N
&=
\left[\begin{array}{ccc}
0 &   M_{L'}^T& 0  \\ 
M_{L'} & 0 & m' \\ 
0  &m'^T & M_{S} \\ 
\end{array}\right],
\end{align}
where $m'\equiv g  v_{H'}/\sqrt2$.
Then, $M_N$ is diagonalized by the unitary matrix $V_N$ as $D_N\equiv V_N^T M_N V_N$ and $N=V_N\psi$, where $D_N$ is mass eigenvalue
and $\psi$ is mass eigenstate.

\begin{figure}[htbp]
   \centering
   \includegraphics[width=5.5cm]{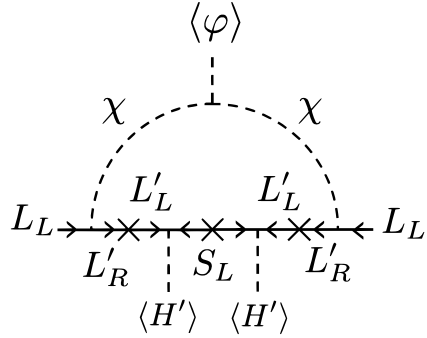} 
   \caption{The diagram which generates the neutrino mass matrix at the one-loop level.  We adopt the flavor eigenstates instead of mass eigenstates to show their connections more clearly. }
   \label{fig:neutrinomass}
\end{figure}
The neutrino mass matrix arises from the following Lagrangian
\begin{align}
-\mathcal{L}_{Y}
&=
 \frac{1}{\sqrt2} \sum_{k=1}^{3} (V_N^\dag)_{a k} f_{k j}\bar\psi_{R_a} \nu_L^j (\chi_R + i \chi_I) +{\rm h.c.},
\label{Eq:lag-mass}
\end{align}
where $\chi\equiv(\chi_R + i \chi_I)/\sqrt2$.
The resulting mass matrix is given through Fig.~\ref{fig:neutrinomass} by
\begin{align}
(m_{\nu})_{ij}
&=\sum_{a=1}^{9}\sum_{k,k'=1}^{3}
\frac{f^T_{ik} (V_N^*)_{ka}D_{N_a} (V_N^\dag)_{ak'}f_{k'j}}{2 (4\pi)^2}
\left[ \frac{m^2_{R}}{m^2_{R}-D^2_{N_a}} \ln \frac{m^2_{R}}{D^2_{N_a}}-
\frac{m^2_{I}}{m^2_{I}-D^2_{N_a}} \ln \frac{m^2_{I}}{D^2_{N_a}}  \right] ~\nn\\
 &=  f^T R f ,\\
R_{kk'}&\equiv  \frac{1}{{2(4\pi)^2}} \sum_{a=1}^{9}(V_N^*)_{ka} { D_{N_a}}
\left[ \frac{m^2_{R}}{m^2_{R}-D^2_{N_a}} \ln \frac{m^2_{R}}{D^2_{N_a}}-
\frac{m^2_{I}}{m^2_{I}-D^2_{N_a}} \ln \frac{m^2_{I}}{D^2_{N_a}} 
 \right] (V_N^\dag)_{ak'},
\end{align}
where $m_{R,I}$ is a mass eigenstate of $\chi_{R,I}$, and
$m_\nu$ is diagonalized by a unitary matrix $U_{\rm PMNS}$~\cite{Maki:1962mu}; $D_\nu\equiv U_{\rm PMNS}^T m_\nu U_{\rm PMNS}$.
Since $R$ is a symmetric three by three matrix, Cholesky decomposition can be done as $R= T^T T$, where $T$ is an upper-right triangle matrix\cite{Nomura:2016run}.
$T$ is uniquely determined by $R$ except their signs, thus we fix all the components of $T$ to be positive.
Then, the Yukawa coupling $f$ is rewritten in terms of the other parameters as follows~\cite{Casas:2001sr}:
\begin{align}
f&= [V_{MNS}^* D_\nu^{1/2} {\cal O} (T^T)^{-1}]^T ,
\end{align}
where ${\cal O}$ is a three by three orthogonal matrix with arbitrary parameters and we take Max[f$_{\alpha i}$]$\lesssim 1.2$
as the perturbative limit.

\subsection{Lepton flavor violations and the anomalous magnetic moment} \label{lfv-lu}
\label{Subsec:LFV}

\begin{figure}[htbp]
   \includegraphics[width=80mm]{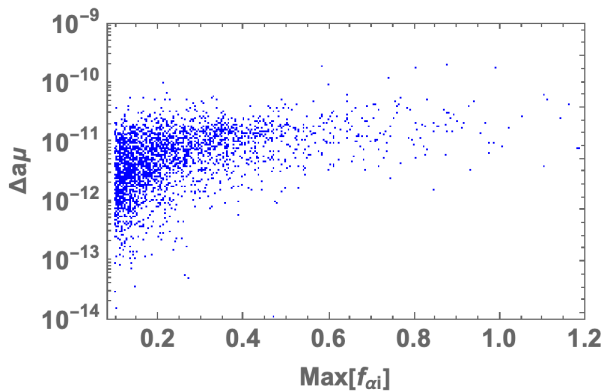}
  \caption{Parameter region in terms of Max[f$_{\alpha i}$] and $\Delta a_\mu$, satisfying the constraints of neutrino oscillation data, LFV processes, and the perturbative limit of  Max[f$_{\alpha i}$]$\lesssim 1.2$.}
  \label{fig.f-damu}
\end{figure}

Before discussing lepton flavor-violations (LFVs),  we define the heavier charged-lepton mass matrix as $D_{E}\equiv V^\dag_{E_L} M_{L'} V_{E_R}$ and $E'^-_{L,R}= V_{E_{L,R}} \psi^-_{L,R}$. Here $D_E$ is mass eigenvalues, and $\psi^\pm_{L,R}$ is mass eigenstates.
Then, LFV processes arise from the following Lagrangian
\begin{align}
{\cal L}_Y
=
\frac{f_{ai}}{\sqrt2} \bar E'_{R_a} \ell_{L_i} (\chi_R + i \chi_I) + \mbox{h.c.}=
{G_{\alpha i}} \bar \psi^-_{R_\alpha} \ell_{L_i} (\chi_R + i \chi_I) + \mbox{h.c.},
\label{eq:lvs-g2}
\end{align}
where $G_{\alpha i}\equiv \sum_{a=1}^3(V^\dag_{E_R})_{\alpha a} f_{ai}/\sqrt{2}$ and $(\ell_1,\ell_2,\ell_3)\equiv (e,\mu,\tau)$~\footnote{The other LFV processes such as $\ell_i\to \ell_j\ell_k\bar\ell_\ell$ and $\mu e\to ee$ have been discussed in refs.~\cite{Nomura:2020azp, Nomura:2020dzw, Toma:2013zsa}.}.
The corresponding branching ratio is given by~\cite{Lindner:2016bgg, Baek:2016kud}
\begin{align}
{\rm BR}(\ell_i\to\ell_j\gamma)
&=
\frac{48\pi^3\alpha_{\rm em} C_{ij}}{(4\pi)^4G_F^2}\left(1+\frac{m^2_{\ell_j}}{m^2_{\ell_i}}\right)
\left|\sum_{\alpha=1}^3\sum_{J=R,I} {G^\dag_{j\alpha} G_{\alpha i} } F(m_J,D_{E_\alpha})\right|^2 ~,\\
F(m_1,m_2)&=\frac{m_2^6 -6 m_2^4 m_1^2 + 3 m_2^2 m_1^4 +2 m_1^6+6 m_2^2 m_1^4\ln\left[\frac{m^2_2}{m^2_1}\right]}{12(m_2^2-m_1^2)^4},
\label{eq:damu1}
\end{align}
where the fine structure constant is $\alpha_{\rm em} \simeq 1/128$, the Fermi constant is $G_F \simeq 1.17\times 10^{-5}$ GeV$^{-2}$, and $(C_{21},C_{31},C_{32}) \simeq (1,0.1784,0.1736)$.
The current experimental upper bounds at 90\% confidence level (CL) are~\cite{TheMEG:2016wtm, Adam:2013mnn}
\begin{align}
{\rm BR}(\mu\to e\gamma) < 4.2\times10^{-13} ~,~
{\rm BR}(\tau\to e\gamma) < 3.3\times10^{-8} ~,~
{\rm BR}(\tau\to \mu\gamma) < 4.4\times10^{-8} ~.
\end{align}
The muon $g-2$ value is positively found via the same interaction with the LFV processes and its form is given by ~\cite{Lindner:2016bgg, Baek:2016kud}
 \begin{align}
 \Delta a_\mu^{(1)} \approx 2 \frac{m_\mu^2}{(4\pi)^2}
  \sum_{\alpha=1}^{3} \sum_{J = R,I} {G^\dag_{2\alpha} G_{\alpha2}} F(m_J,D_{E_\alpha}),
\label{amu1L}
 \end{align}
 where the discrepancy of the muon $g-2$ value, between the experimental measurement and the SM prediction, is given by~\cite{Hagiwara:2011af} 
\begin{align}
\Delta a_{\mu}=(26.1 \pm 8.0)\times 10^{-10}. 
\end{align}
We preform a global numerical analysis that satisfies the constraints of neutrino oscillation data, LFV processes, and the perturbative limit of Max[f$_{\alpha i}$]$\lesssim 1.2$. In Fig.~\ref{fig.f-damu}, we illustrate the allowed region between Max[f$_{\alpha i}$] and $\Delta a_\mu$, from which the maximum value of $\Delta a_\mu$ is about $2\times 10^{-10}$. Even though it does not reach the measured value of $\Delta a_\mu\sim 10^{-9}$, we expect it can be tested by future experiments such as Belle II~\cite{Batell:2017kty} soon.

\section{Dark matter}
This model is also interesting in that it includes neutral fields $\chi$, the lightest particle of $N_R$, and $\psi$; all which can be DM candidates. In terms of the energy density, in units of the critical density, they are constrained by \cite{Aghanim:2018eyx}
\begin{align}
\Omega_{\mathrm{DM}} h^2=0.120 \pm 0.001,
\label{eq:omegaDM}
\end{align}
where $\Omega_{\mathrm{DM}}$ and $h$ are energy density of DM in unit of the critical density and the Hubble parameter, respectively.
Among the electrically neutral fields described in Table \ref{tab:1},   the lightest neutral fermion $N_R$ or $\psi$, and the inert boson $\chi$ are stable because of the $Z_2\times Z'_2$ parity.  Recall that $Z'_2$ is a residual symmetry after the spontaneous symmetry breaking of $U(1)_R$.
%

If $N_R$ is the main component of the lightest $Z_2$ odd particle, it can be a dark matter candidate; however, this case is not interesting because the relic abundance is independent of the parameter space shown in Fig. \ref{fig.f-damu}.
On the other hand, because $\psi$ and $\chi$ interact with each other via the coupling $f_{\alpha i}$ that is constrained in Fig. \ref{fig.f-damu}. If $m_\chi > m_\psi $ and the coupling $f_{\alpha i}$ is not small, only $\psi$ can be dark matter since all the $\chi$ particles decay into $\psi$ through interactions. In this instance; unfortunately, we cannot expect $\psi$'s detection in direct detection experiments for the near future, because it does not interact with quarks directly. 
Therefore, we will focus on $\chi$ as the dark matter candidate in our model, and investigate the parameter space allowed by both neutrino physics and dark matter constraints.

Before estimating the features of $\chi$ as DM, we check the parameter space that the lightest one of $N_R$ cannot be the dominant component of DM. If the lightest one of $N_{R_1}$; $X_R$, is to be DM, its main annihilation cross section 
for the relic abundance calculation
comes from s-channels via $\varphi$ and $Z'$, and their forms are given up to the s-wave by 
\begin{align}
(\sigma v)_{X_{R_1}\bar X_{R_1} \to 2h} &\approx
\frac{y_{N_{11}}\mu_{\varphi hh}^2}{512\pi(m_{\varphi}^2-m_X^2)^2}\sqrt{1-\frac{m_h^2}{m_X^2}},\label{eq:nr-h}\\
(\sigma v)_{X_{R_1}\bar X_{R_1} \to f \bar f} &\approx
\sum_{f=u_R,d_R,e_R}
\frac{(g'x)^4 m^2_f}{32\pi(4 m_X^2 - m_{Z'}^2)^2}
\sqrt{1-\frac{m_f^2}{m_X^2}},\
 (f=u_R,d_R,e_R), \label{eq:nr-zp}
\end{align}
where $h$ is the SM Higgs mainly originating from $H$, $m_\varphi$ is the mass eigenstate of the $\varphi$'s CP even component, and $\mu_{\varphi hh}$ is a trilinear coupling between $\varphi$ and $h$ that is described by the linear combinations of parameters in the Higgs potential. We neglect any mixings among any bosons as well as the decay rate of $\varphi$.
Here, we estimate these cross sections by fixing the typical parameters.
If we fix the parameters to be $m_X=500$ GeV, $m_\varphi=700$ GeV, $y_{N_{11}}=1$, $m_h=125$ GeV in Eq.~(\ref{eq:nr-h}),
and we restrict the cross section to be larger than 
4 pb
for each of the modes,~\footnote{Typical scale to explain the relic density is about 
0.4 pb.} then we find the following bound:
\begin{align}
 2150\ {\rm GeV} \lesssim \mu_{\varphi hh} .
\end{align}
In the case of Eq.~(\ref{eq:nr-zp}), we find the following bound
\begin{align}
 1.58 \lesssim g'x,
\end{align}
where we apply the relation $g'/m_{Z'}\le1/$(3.7 TeV) obtained by LEP experiment~\cite{Schael:2013ita}.
As long as these constrains are satisfied, direct detections for $X_{R_1}$ are allowed by the current experimental bound~\cite{Kanemura:2010sh}.


As well as the abundance of DM, its direct detection places a severe constraint on the DM-nucleon interaction. The most stringent constraint is given by the XENON1T experiment, which excludes the spin-independent DM-proton cross section 
\begin{align}
\sigma_{p}^\textrm{SI} \gtrsim 10^{-9} \textrm{pb},
\label{eq:DDSI}
\end{align}
for $O(100)$ GeV DM mass \cite{Aprile:2018dbl}. 

In our model, the dominant spin-independent elastic scattering cross section arises from the process via $Q'$ exchange.
 Its effective Lagrangian at the component level is given by
\begin{align}
{\cal L} \simeq - i \sum_{i=1}^{3}  \sum_{\alpha =1}^3 \frac{h^\dag_{i\alpha} h_{\alpha i}}{2 M^2_{Q'_\alpha}}
[\bar q_i \gamma^\mu P_L q_i ][\chi^* \overleftrightarrow\partial_\mu\chi],
\end{align}
where we have assumed the four transferred momentum squared $q^2$ is much smaller than $M^2_{Q'_\alpha}$ and the nucleus of the target almost stops (is at its rest frame).
We then straightforwardly define the DM-nucleon elastic scattering operator as follows:
\begin{align}
{\cal L}_N &\simeq  - i  \sum_{i=u}^{d,s} \sum_{\alpha =1}^3 \frac{h^\dag_{i\alpha} h_{\alpha i}}{2 M^2_{Q'_\alpha}}
 [ F^{q_i/N}_1\bar N \gamma^\mu N
  - F^{q_i/N}_A\bar N \gamma^\mu \gamma_5 N ] (\chi^* \overleftrightarrow\partial_\mu\chi ),
\end{align} 
where $F^{q_i/N}_{1,A}$ are the form factors given by Ref.~\cite{Bishara:2017pfq}.
Then, the squared matrix element is
\begin{align}
  \label{eq:infer}
|{\cal M}|^2 &
\simeq
2 m_N^2 m_\chi^2  \sum_{i=1}^{3}  \sum_{\alpha=1}^3
\frac{ |h^\dag_{i\alpha} h_{\alpha i}|^2}{M^4_{Q'_\alpha}}
\left(
\left|F_1^{q_i/N}\right|^2 + \left| F_A^{q_i/N}\right|^2
\right),
\end{align}
where $F_A^{q_i/N}$ contributes only to the spin dependent cross section that are much less constrained than the spin independent one, thus we neglect this term.~\footnote{We would like thank the referee for pointing it out.}
%
Finally, the complete form of the DM-nucleon elastic scattering cross section is expressed by~\cite{Hutauruk:2019crc}
\begin{align}
\sigma^{SI} &\approx \left(\frac{m_\chi m_N}{m_N+m_\chi}\right)^2  \frac{|{\cal M}|^2}{32\pi m_\chi^2m_N^2}
=
\frac1{16\pi }
\left(\frac{m_\chi m_N}{m_N+m_\chi}\right)^2
\sum_{i=1}^{3}  \sum_{\alpha=1}^3
\frac{ |h^\dag_{i\alpha} h_{\alpha i}|^2}{M^4_{Q'_\alpha}}\left|F_1^{q_i/N}\right|^2,
\label{eqq:dd}
\end{align}
where $\sum F_1 \approx 3$ corresponds to the effective operator
$(\bar N \gamma^\mu N )(\chi^* \overleftrightarrow{\partial_\mu}\chi)$, 
$\sum F_A\approx 0.49$ corresponds to the effective operator
$(\bar N \gamma^\mu\gamma_5 N)(\chi^*\overleftrightarrow{\partial_\mu}\chi)$,
and $m_N \approx 0.939$ GeV.
%

In Section \ref{sec:NA}, we investigate parameter region satisfying the constraints of the relic abundance and the direct detection.

\subsection{Numerical analysis \label{sec:NA}}
\begin{figure}[htbp]
   \includegraphics[width=80mm]{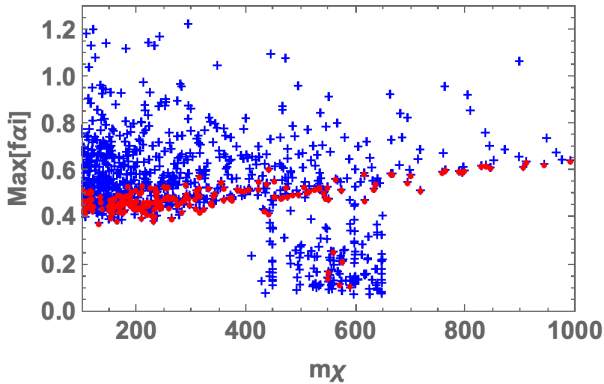}
  \caption{Parameter region satisfying the constraints of the relic abundance of DM and direct searches. Red \textcolor{red}{$\bullet$} and blue \textcolor{blue}{$+$} represent points for constraint to agree with Eq. (\ref{eq:omegaDM}) within 3$\sigma$ and $\Omega_\mathrm{DM}h^2 \leq 0.120$, respectively. All the points satisfy the constraints by direct search Eq.~(\ref{eq:DDSI}) and LFVs described in Fig.\ref{fig.f-damu}.} 
  \label{fig:DM}
\end{figure}
We show in Fig.\ref{fig:DM}, the parameter region satisfying the constraints of the thermal relic abundance, the direct detection of DM, as well as the constraints that are discussed in Sec.\ref{lfv-lu}. Random scans of the regions for the parameters $900\leq m_{Q'^{a}}, m_{N_R^a}, m_{S_L^a}, m_{L'}, m_\varphi \leq 1300$ GeV, $0.002 \leq h_{\alpha i} \leq 0.01$, $0.002 \leq $ (couplings of the Higgs potential in Eq.(\ref{Eq:pot})) $ \leq 0.01$ are performed.
The thermal relic abundance can be obtained by solving a Boltzmann equation, and is approximately represented as 
$\Omega_{\mathrm{DM}} h^2\simeq 0.05\mathrm{[pb]}/\langle \sigma v\rangle$, where $\langle \sigma v \rangle$ represents the thermal average of DM that is constructed by the annihilation cross section $\sigma v$ in the non-relativistic limit.~\footnote{Due to rather complicated process to compute relic abundance via the thermal cross section, we calculate it  with Micromegas 5.0 \cite{Belanger:2018ccd}.}
In the bulleted list below, we highlight the nature of DM in our model and its relation to the muon g-2 value and the dependence on the valid input parameters.

\begin{itemize}
\item In the region where Max[$f_{\alpha i}$] is larger than 0.4, the annihilation process $\chi \chi \to \tau' \to \tau \bar{\tau}$ and $\chi \chi \to \mu' \to \mu \bar{\mu}$ dominate the contribution to the relic abundance due to a relatively large $f_{\alpha i}$ coupling. {The annihilation cross section is approximately represented as 
\begin{align}
(\sigma v)_{\chi \chi \to \tau \bar{\tau}}= \frac{m_\chi^2 |f_{i\alpha}|^4}{48\pi(m_{\tau'}^2+m_\chi^2)^2}  v^2.
\end{align}
The $\chi \chi \to  \mu \bar{\mu}$ annihilation process only $\tau\to\mu$ is replaced. Here we have expanded the cross section in terms of the relative velocity $v$.}
The scattering process q$\chi\to Q'\to$q$\chi$ contributes to the direct detection and it gives constraint to the coupling $h_{\alpha i}$, typically of $h_{\alpha i} \leq 0.008$ from Eqs.~(\ref{eq:DDSI}) and~(\ref{eqq:dd}).
\item In the region $450 < m_\chi < 650$ GeV and when Max[$f_{\alpha i}$] is smaller than 0.4, there are also points satisfying the constraints. Here the annihilation process $\chi \chi \to \varphi \to h h$ is the main contribution to the thermal production. 
The annihilation cross section is approximately given as,
\begin{align}
(\sigma v)_{\chi \chi \to h h} \simeq \frac{\lambda_{\varphi H}^2\mu_1^2 v_\varphi^2 \sqrt{m_\chi^2-m_H^2}}{512 m_\chi^3 (4 m_\chi^2-m_\varphi^2)^2}.
\end{align}
Since the $\varphi$ mass is taken as $900\leq  m_\varphi \leq 1300$ GeV, the annihilation cross section is enhanced due to $\varphi$ resonance in this region. 
With a smaller $f_{\alpha i}$ than that, coannihilation processes including $N_R^a$ can be enhanced if $\chi$ and one of the $N_R^a$ have degenerate masses. However, the parameter region such that the coannihilation processes give a dominant contribution is not favored, because it is difficult to satisfy both the relic abundance and the direct detection constraints.
\item Taking the result of the DM analysis into consideration; 0.4 $\lesssim$ Max$[f_{\alpha i}] \lesssim$ 0.6,
we find that muon g-2 value is, at most, about $10^{-10}$.
\end{itemize}

\section{Semi-leptonic decays and neutral meson mixings}
Now we discuss the semi-leptonic decays of $b\to s\bar\ell\ell$. If the DM from our model is definitely a real scalar these processes would vanish, but if the DM is complex we find non-vanishing contributions from box diagrams.
Furthermore, we can approximately consider it to be a complex scalar, since the mass difference between $\chi_R$ and $\chi_I$ is expected to be very small to get larger Yukawa couplings of $f$. 
In our model, we have the following effective Hamiltonian at the box one-loop diagram~\cite{Arnan:2016cpy, Chiang:2017zkh}:
\begin{align}
{\cal H}_{eff}&=\sum_{a,b=1}^3 h^\dag_{3a}h_{a2} G^\dag_{\ell b}G_{b\ell} F(M_{Q'_a},D_{E_b})(\bar s\gamma_\mu P_L b - \ell\gamma P_L \ell)\nn\\
&\equiv -C_{SM}[C^{\ell\ell}_{9} {\cal O}_9 - C^{\ell\ell}_{10} {\cal O}_{10}],\\
F(m_a,m_b)&=\frac{1}{2(4\pi)^2}\int_0^1 [dx_3]\frac{xdx}{x m_\chi^2 + y M^2_{Q'_a} + z D^2_{E_a} },\quad
C_{SM}\equiv \frac{V_{tb} V^*_{ts} G_F \alpha_{em}}{\sqrt2 \pi},
\end{align}
where $[dx_3]\equiv dxdydz\delta(1-x-y-z)$, $G_F\equiv 1.17\times 10^{-5}$ GeV$^{-2}$ is Fermi constant, $\alpha_{em}\equiv 1/128.9$ is the electromagnetic fine structure constant, and $V_{tb}\text{ and }V_{ts}$ are the 3-3 and 3-2 components of the Cabbibo-Kobayashi-Maskawa (CKM) matrix, respectively. Therefore, the new contribution to the $C_9^{\ell\ell}$, under the condition of $C_9^{\ell\ell}=-C_{10}^{\ell\ell}$, is given by
\begin{align}
C^{\ell\ell}_9= -\frac{1}{C_{SM}} \sum_{a,b=1}^3 h^\dag_{3a}h_{a2} G^\dag_{\ell b}G_{b\ell} F(M_{Q'_a},D_{E_b}).
\end{align}
In instance of $\ell=\mu$ an anomaly is reported, and its deviation is given by [-0.85,-0.50]([-1.22,-0.18]) at 1$\sigma$(3$\sigma$) CL with the best fit value of -0.68~\cite{Descotes-Genon:2015uva}.
When we preform these calculations we have to consider the process of $B_{d/s}\to \bar\mu\mu$ that gives the following constraint~\cite{Sahoo:2015wya}:
\begin{align}
C^{\mu\mu}_{31(2)}\equiv\sum_{a,b=1}^3 h^\dag_{3a}h_{a1(2)} G^\dag_{2 b}G_{b2} F(M_{Q'_a},D_{E_b})\lesssim 5(3.9)\times 10^{-9} {\rm GeV}^{-2}.
\end{align}

The neutral meson mixings also give constraints~\cite{Gabbiani:1996hi}. The most stringent constraint is  $B_s-\bar B_s$ mixing, which is denoted by $\Delta m_{B_s}$, and our new contribution is restricted by
\begin{align}
\Delta m_{B_s}\approx \frac{2}{3} m_{B_s} f^2_{B_s} \sum_{a,b=1}^3 h^\dag_{3a}h_{a2} h^\dag_{3b} h_{b2}  F(M_{Q'_a},M_{Q'_b}),
\end{align}
where $f_{B_s}\approx 0.274$ GeV $m_{B_s}\approx 5.367$ GeV~\cite{DiLuzio:2017fdq, DiLuzio:2018wch}.
Then, we find the following constraint~\cite{Hutauruk:2019crc, Kumar:2020web}:
\begin{align}-2.27\times 10^{-12}{\rm GeV}\lesssim \Delta m_{B_s} \lesssim 1.07\times 10^{-12} {\rm GeV}.\end{align}

Taking the result of our preceding numerical analysis into account, we estimate these values.
Then, we found that all of them are tiny compared to observables.
Therefore, their maximum values are given by 
\begin{align}
&{\rm Max}[C^{\mu\mu}_{9}]\approx 2.43\times 10^{-7},\\
& {\rm Max}[C^{\mu\mu}_{31}]\approx 3.95\times 10^{-16},\ {\rm Max}[C^{\mu\mu}_{32}]\approx 3.96\times 10^{-16},\\
& \Delta m_{B_s}=3.52\times 10^{-22}{\rm GeV}.
\end{align}
Even though we cannot explain the $b\to s\mu\bar\mu$ anomaly,
the other constraints are allowed in our model.
Moreover, there is a possibility to explain the DM and the $b\to s\mu\bar\mu$ anomaly simultaneously, if we consider the fermionic DM candidate; the lightest particle of $N_R$ or $\psi$.
Because the fermionic DM does not couple to any quarks, $h$ is not restricted by the direct detection of DM searches. Thus, a large $h$ is allowed. In the instance of the lightest $N_R$ DM, it couples neither to $f_{\alpha i}$ nor to $h_{\alpha i}$, which are then free from the constraints of DM origins. Thus, we have more degrees of freedom available.
A detailed analysis will be preformed in a future work.~\footnote{Similar discussions have been done by refs.~\cite{Borah:2020swo, Barman:2018jhz}, in  which they seriously evaluate semi-leptonic decays and meson mixings, too.}


\section{Summary and Conclusions}
 We have studied the neutrino and WIMP-DM based on a gauged $U(1)_R$ symmetry in the framework of a radiative seesaw scenario.
 The bosonic DM can interact with quark and lepton sectors through vector-like heavier quark and leptons, and plays a role in generating neutrino mass matrix together with neutral heavier fermions.
 We have found several constraints for related masses and couplings on DM by comparing our theoretical results with the experimental ones of relic density and the direct detection searches.
Considering all the constraints, we have obtained that the maximum muon g-2 value is of the order $10^{-10}$ and might be tested in future experiments.


\begin{acknowledgements}
KIN was supported by  JSPS  Grant-in-Aid  for  ScientificResearch (A) 18H03699, (C) 21K03562, 21K03583, and Wesco research grant.
This research was supported by an appointment to the JRG Program at the
APCTP through the Science and Technology Promotion Fund and Lottery Fund
of the Korean Government. This was also supported by the Korean Local
Governments - Gyeongsangbuk-do Province and Pohang City (H.O.).
H.O.~is sincerely grateful for the KIAS member. 
\end{acknowledgements}

\end{document}